\documentclass[manuscript]{aastex}

\slugcomment{Not to appear in Nonlearned J., 45.}

\shorttitle{Subaru Transitional disk around Upper Sco}
\shortauthors{Mayama et al.}

\begin{document}

\title{SUBARU IMAGING OF ASYMMETRIC FEATURES IN A\\
TRANSITIONAL DISK IN UPPER SCORPIUS
\footnote{
    Based on data collected at the Subaru Telescope, which is operated
    by the National Astronomical Observatory of Japan.}
    }

\author{S. Mayama\altaffilmark{1,2},
J. Hashimoto\altaffilmark{3},
T. Muto\altaffilmark{4},
T. Tsukagoshi\altaffilmark{5},
N. Kusakabe\altaffilmark{3},
M. Kuzuhara\altaffilmark{3,6},
Y. Takahashi\altaffilmark{3,7},
T. Kudo\altaffilmark{8},
R. Dong\altaffilmark{9},
M. Fukagawa\altaffilmark{10},
M. Takami\altaffilmark{11},
M. Momose\altaffilmark{5},
J. P. Wisniewski\altaffilmark{27},
K. Follette\altaffilmark{15},
L. Abe\altaffilmark{12},
E. Akiyama\altaffilmark{3},
W. Brandner\altaffilmark{13},
T. Brandt\altaffilmark{9},
J. Carson\altaffilmark{14},
S. Egner\altaffilmark{8},
M. Feldt\altaffilmark{13},
M. Goto\altaffilmark{29},
C. A. Grady\altaffilmark{16,17,18},
O. Guyon\altaffilmark{8},
Y. Hayano\altaffilmark{2,8},
M. Hayashi\altaffilmark{2,3},
S. Hayashi\altaffilmark{2,8},
T. Henning\altaffilmark{13},
K. W. Hodapp\altaffilmark{19},
M. Ishii\altaffilmark{8},
M. Iye\altaffilmark{2,3},
M. Janson\altaffilmark{9},
R. Kandori\altaffilmark{3},
J. Kwon\altaffilmark{2},
G. R. Knapp\altaffilmark{9},
T. Matsuo\altaffilmark{20},
M. W. McElwain\altaffilmark{18},
S. Miyama\altaffilmark{21},
J.-I. Morino\altaffilmark{3},
A. Moro-Martin\altaffilmark{9,22},
T. Nishimura\altaffilmark{8},
T.-S. Pyo\altaffilmark{8},
E. Serabyn\altaffilmark{23},
H. Suto\altaffilmark{3},
R. Suzuki\altaffilmark{3},
N. Takato\altaffilmark{2,8},
H. Terada\altaffilmark{8},
C. Thalmann\altaffilmark{24},
D. Tomono\altaffilmark{8},
E. L. Turner\altaffilmark{9,25},
M. Watanabe\altaffilmark{26},
T. Yamada\altaffilmark{28},
H. Takami\altaffilmark{2,8},
T. Usuda\altaffilmark{2,8},
M. Tamura\altaffilmark{2,3}}

\altaffiltext{1} {The Center for the Promotion of Integrated Sciences, The Graduate University for Advanced Studies~(SOKENDAI), Shonan International Village, Hayama-cho, Miura-gun, Kanagawa 240-0193, Japan}
\email{mayama\_satoshi@soken.ac.jp}

\altaffiltext {2} {Department of Astronomical Science, 
The Graduate University for Advanced Studies~(SOKENDAI), 
2-21-1 Osawa, Mitaka, Tokyo 181-8588, Japan}
\altaffiltext {3} {National Astronomical Observatory of Japan, 2-21-1
Osawa, Mitaka, Tokyo 181-8588, Japan}
\altaffiltext{4} {Division of Liberal Arts, Kogakuin
University, 1-24-2, Nishi-Shinjuku, Shinjuku-ku, Tokyo, 163-8677, Japan}
\altaffiltext {5} {College of Science, Ibaraki University, 2-1-1
Bunkyo, Mito, Ibaraki 310-8512, Japan }
\altaffiltext{6} {Department of Earth and Planetary Science, The University of Tokyo, Hongo 7-3-1, Bunkyo-ku, Tokyo 113-0033, Japan}
\altaffiltext{7} {Department of Astronomy, The University of Tokyo,
Hongo 7-3-1, Bunkyo-ku, Tokyo 113-0033, Japan}
\altaffiltext{8} {Subaru Telescope, 650 North A'ohoku Place, Hilo, HI
96720, USA}
\altaffiltext {9} {Department of Astrophysical Sciences,
Princeton University, NJ08544, USA}
\altaffiltext {10} {Department of Earth and Space Science, Graduate
School of Science, Osaka University, 1-1, Machikaneyama, Toyonaka, Osaka
560-0043, Japan}
\altaffiltext {11} {Institute of Astronomy and Astrophysics, Academia
Sinica, P.O. Box 23-141, Taipei 106, Taiwan}
\altaffiltext{12} {Laboratoire Hippolyte Fizeau, UMR6525, Universite de
Nice Sophia-Antipolis, 28, avenue Valrose, 06108 Nice Cedex 02, France}
\altaffiltext{13} {Max Planck Institute for Astronomy, Koenigstuhl 17, 69117 Heidelberg, Germany}
\altaffiltext{14} {Department of Physics and Astronomy, College of Charleston, 58 Coming St., Charleston, SC 29424, USA.}
\altaffiltext {15} {Department of Astronomy and Steward Observatory, 
The University of Arizona, 933 North Cherry Avenue, Rm. N204, Tucson, AZ 85721-0065, USA}
\altaffiltext {16} {Goddard Center for Astrobiology, NASA's Goddard Space Flight Center, Greenbelt, MD 20771, USA}
\altaffiltext {17} {Eureka Scientific, 2452 Delmer, Suite 100, Oakland CA
96002, USA}
\altaffiltext {18} {ExoPlanets and Stellar Astrophysics Laboratory, Code
667, Goddard Space Flight Center, Greenbelt, MD 20771 USA}
\altaffiltext{19} {Institute for Astronomy, University of Hawaii, 640
North A'ohoku Place, Hilo, HI 96720, USA}
\altaffiltext{20} {Department of Astronomy, Kyoto University,
Kitashirakawa-Oiwake-cho, Sakyo-ku, Kyoto, 606-8502, Japan}
\altaffiltext {21} {Office of the President, Hiroshima University,
1-3-2 Kagamiyama, Higashi-Hiroshima, Hiroshima 739-8511, JAPAN}
\altaffiltext{22} {Departamento de Astrofisica, CAB (INTA-CSIC),
Instituto Nacional de T\'ecnica Aeroespacial, Torrej\'on de Ardoz,
28850, Madrid, Spain}
\altaffiltext{23} {Jet Propulsion Laboratory, California Institute of
Technology, Pasadena, CA 91109, USA}
\altaffiltext{24} {Astronomical Institute ''Anton Pannekoek", University
of Amsterdam, Science Park 904, 1098 XH Amsterdam, The Netherlands}
\altaffiltext{25} {Institute for the Physics and Mathematics of the
Universe, The University of Tokyo, Kashiwa 227-8568, Japan}
\altaffiltext{26} {Department of Cosmosciences, Hokkaido University,
Sapporo 060-0810, Japan}
\altaffiltext{27} {H L Dodge Department of Physics \& Astronomy, University of Oklahoma, 440 W Brooks St. Norman, OK 73019, USA}
\altaffiltext{28} {Astronomical Institute, Tohoku University, Aoba,
Sendai 980-8578, Japan}
\altaffiltext{29} {Universit\"ats-Sternwarte M\"unchen Scheinerstr. 1, D-81679 Munich, Germany}


\begin{abstract}
We report high-resolution~(0.07 arcsec) near-infrared polarized intensity
 images of the circumstellar disk around the star 2MASS J16042165-2130284 obtained with HiCIAO mounted on the Subaru 8.2 m
 telescope.  We present our $H$-band data, which clearly exhibits a 
resolved, face-on disk with a large inner hole for the first time at
 infrared wavelengths.  We detect the centrosymmetric polarization pattern
in the circumstellar material as has been observed in other disks.  Elliptical
 fitting gives the semimajor axis, semiminor axis, and 
 position angle (P.A.) of the disk as 63 AU, 62 AU, and -14 $^{\circ}$, respectively.
 The disk is asymmetric, with one dip located at P.A.s of $\sim$85$^{\circ}$.
 Our observed disk size agrees well with a previous study of dust and CO emission at submillimeter wavelength with Submillimeter Array.  Hence, the near-infrared light is interpreted as scattered light reflected
 from the inner edge of the disk.  Our observations also detect an
 elongated arc (50 AU) extending over the disk inner hole.  It emanates at the inner edge of the western side of the disk, extending inward first, then curving to the northeast.
We discuss the possibility that the inner hole, the dip, and the arc that we have observed may be related to the existence of unseen bodies within the disk.
\end{abstract}



\section{Introduction}

Planets are believed to form in protoplanetary disks.  To
study the structures of protoplanetary disks, multi-wavelength observations have targeted such disks around young stellar objects located in nearby star
forming regions such as the Taurus, $\rho$ Oph, and Upper Scorpius regions
 \citep[e.g.,][]{1995ApJS..101..117K,1999AJ....117.2381P,2006ApJS..165..568F,2009ApJ...700.1502A, 2009AJ....137.4024D,2009ApJ...693L..81M, 2011ApJ...732...42A,2012ApJ...745...23M}. 
Among such nearby star forming regions, Upper Scorpius, which is located
at a distance of 145 pc \citep{1999AJ....117..354D}, provides a unique
environment, particularly in terms of its age.  While the age is only 
$\sim$5--10~Myr\citep{2012ApJ...746..154P, 2012AJ....144....8S},
 the lack of dense molecular material and embedded young
stellar objects indicate that the star formation process in Upper
Scorpius is over. The area is essentially free of dense gas and dust
clouds, and the association members show only moderate extinctions
(A$_\mathrm{V} <$ 2 mag).  In fact, there are only two $K$ and $M$
stars (1.5\%), including the star discussed here, in the Upper Scorpius region which exhibit a $K$-band
excess, although about half of the stars in Taurus exhibit a $K$-band
excess \citep{1989AJ.....97.1451S}.  This implies that there are very
few stars still hosting protoplanetary disks in Upper Scorpius star forming region.  

Here, we introduce our target 2MASS J16042165-2130284 (hereafter
J1604-2130) in Upper Scorpius.  J1604-2130 was first identified as a
member of the Upper Scorpius OB association in the spectroscopic survey
of X-ray selected sources of \citet{1999AJ....117.2381P},
 who found: stellar age 3.7~Myr, spectral type K2, stellar mass 1~$M$$_\odot$, 
$\log T_\mathrm{eff}$[K]=3.658, and $\log[L/L~_\odot]=-0.118$.
The spectral energy distribution (SED) of J1604-2130 has the features of a transitional disk \citep{2009AJ....137.4024D}, in which small dust grains have been partly cleared from the inner disk but are still abundant at
larger radii.  Recently, \citet{2012ApJ...753...59M} presented the Sub-Millimeter Array (SMA) 880 ~$\mu$m images, which directly resolved a face-on protoplanetary disk around J1604-2130, and derived a dust mass of 0.1~$M$$_\mathrm{Jup}$.
The availability of radio data was one advantage to selecting this target, however other advantages abound and are as follows:
1. J1604-2130 is located in a nearby star forming
region, Upper Sco; 2. it hosts one of the largest inner holes in transitional
disks discovered to date; 3. it hosts the most massive disk in Upper
Scorpius; 4. the inclination $6^{\circ}\pm1.5^{\circ}$ \citep{2012ApJ...753...59M} is very close to face-on.

Infrared photometry and spectroscopy with Spitzer has shown that
less than $\sim$10\% of stars with disks show some degree of inner disk depletion yet still retain massive outer disks \citep[e.g.,][]{2010ApJ...712..925C,2010ApJ...708.1107M,2011ApJ...730L...9L}.
Although there have been several transitional disks detected
with radio observations with SMA around single young solar mass sources
\citep[e.g.,][]{2006A&A...460L..43P,2011ApJ...732...42A},
only a handful of transitional disks have been detected in the near-infrared~\citep[e.g.,][]{2006ApJ...636L.153F,2010ApJ...718L..87T,2011ApJ...729L..17H}. 
This is mainly because it is difficult to resolve structures with 
only few 10 AU scale close to a bright central star.
Furthermore, structures inside the hole of disks have rarely been directly imaged or resolved to date.  To gain a better
understanding of the morphological structures at radii where planets are expected to form, we conducted high-resolution
near-infrared polarimetric observations of J1604-2130.  

\section{Observations and Data Reduction}

We carried out polarimetry in $H$-band (1.6~$\mu$m) toward J1604-2130 
using the high resolution imaging instrument HiCIAO 
\citep{2006SPIE.6269E..28T,2006SPIE.6269E.123H} with a dual-beam  
polarimeter at the Subaru 8.2~m Telescope on 2012 April 11. These observations are part of the ongoing high-contrast imaging survey
\citep[SEEDS;][]{2009AIPC.1158...11T}. The polarimetric observation mode
acquires \textit{o}-rays 
and \textit{e}-rays simultaneously, and images a field of view of
10$''\times$20$''$ with a pixel  
scale of 9.53~mas/pixel. J1604-2130 was observed without an occulting mask in order to image 
the inner most region around the central star. The exposures were performed at four position 
angles (P.A.s) of the half-wave plate, with a sequence of P.A. = 0$^{\circ}$, 45$^{\circ}$, 
22.5$^{\circ}$, and 67.5$^{\circ}$ to measure the Stokes parameters. The integration time 
per wave plate position was 15 sec and we obtained twenty-five waveplate cycles.
The adaptive optics system \citep[AO188;][]{2010SPIE.7736E..21H} provides a diffraction  
limited and mostly stable stellar PSF.  The total integration time of
the polarization intensity (hereafter $PI$ )  
image was 750 sec after removing low quality images with large FWHMs by
careful inspections of the stellar PSF.

The Image Reduction and Analysis Facility (IRAF\footnote{
    IRAF is distributed by National Optical Astronomy Observatory, which is operated by the Association of Universities for Research in Astronomy, Inc., under cooperative agreement with the National Science Foundation.}) software was used for
all data reduction.   
The polarimetric data reduction technique is described by
\citet{2011ApJ...729L..17H} and \citet{2012ApJ...748L..22M},  
using the standard approach for polarimetric differential imaging (Hinkley et al. 2009).  
By subtracting two images of extraordinary and ordinary rays at each wave plate position, 
we obtained $+Q$, $-Q$, $+U$, and $-U$ images, from which $2Q$ and $2U$ images were made 
by another subtraction to eliminate remaining aberration. The $PI$ was then
calculated by $PI =\sqrt{Q^2+U^2}$.  
The instrumental polarization of HiCIAO at the Nasmyth platform was
corrected for by following \citet{2008SPIE.7016E..48J}.


\section{Results}
The $H$-band $PI$ image of J1604-2130 is presented in Figure 1(a).  
The emission arises from 
dust particles mixed with gas in the circumstellar structures, scattering the stellar light.  
Details of the near-infrared disk around J1604-2130 have 
for the first time emerged at 0.07 arcsec.  The disk has a dip D, located around a position angle $\sim$85$^{\circ}$, making this transitional disk asymmetric.
Inside the bright disk, we see a region where the surface
brightness drops by a factor of $\sim$5 compared with the disk.
We call this inner lower-brightness region the ``hole'' throughout the remainder of this paper.
Figure 1(b) shows $H$-band polarization vectors.
The disk with its centrosymmetric polarization pattern
 surrounds the central source J1604-2130 as previously observed in other disks using the technique of near-infrared imaging polarimetry~\citep[e.g.,][]{2011ApJ...729L..17H}.

 Elliptical fitting was performed to measure several disk parameters including 
 the angular separations of the major and minor axes, the position angle, inclination, 
and the position of the ellipse center.  We first measured disk radial profiles in 5$^{\circ}$ position angle increments, and then extracted coordinates of the brightest peak area.  
 Those coordinates were used to fit an ellipse.  The nonlinear least-squares 
 Levenburg-Marquardt algorithm with the above five free parameters was performed 
 and its result is displayed in Table 1. Its shape is superimposed on Figure 1(a).    The derived semimajor and semiminor axes are called ``NIR peak radii'' in this paper. 
 The geometric center of the disk is consistent with the position of the star within errors.
 The disk position angle has a relatively large error because the disk is
 close to circular.

 Figure \ref{fig3} shows the azimuthal surface brightness at $r$ = 33, 63, 145~AU, which are illustrated in figure 1(c), with the position angle measured from north to east.  The error bars shown in figure \ref{fig3} represent the calculated standard deviation.  The left panel of figure \ref{fig4} shows the radial surface brightness profile of J1604-2130 along the minor axis.
 The error bars shown in figure \ref{fig4} represent the calculated standard deviation.
 In the north, the surface brightness along the minor axis increases as $r$$^{3.6\pm0.2}$ from 40 to 62~AU, and decreases as $r$$^{-4.7\pm0.1}$ from 74 to 124~AU. In the south, the surface brightness along the minor axis increases 
as $r$$^{3.0\pm0.4}$ from 40 to 62~AU, and decreases 
as $r$$^{-4.7\pm0.2}$ from 74 to 124~AU.  The right panel of Figure \ref{fig4}
shows the radial profile of the surface brightness along
the major axis. 
In the east, the surface brightness along the major axis increases as
$r$$^{2.6\pm0.2}$ from 40 to 62~AU, and decreases as $r$$^{-4.7\pm 0.1}$ from 74 to
124 AU. In the west, the surface brightness along the major axis increases as
$r$$^{1.9\pm0.2}$ from 40 to 62 AU, and decreases as $r$$^{-4.0\pm0.2}$ from 74 to
124 AU.  The profiles along both the major and minor axes show a change of
slope beyond 63 AU, thus reconfirming the NIR peak radius to be 63~AU.  Along all four axes, the counts drop to half of the peak counts at around 50~AU from the central star.
The width of the brightest region where counts exceeds half of the peak counts in the disk is around 30 AU.

A salient feature inside the hole is an elongated arc extending over the disk inner hole.  It begins at the inner edge of the western side of the disk, extending inward first, then curving to the northeast.  The whole arc extends at least 50~AU.
This arc is equivalent in brightness to the inner edge of the disk and to the dip D region.  This arc consists of a real feature and is not an artifact because in contrast to the inner hole errors (shown as squares in Figure \ref{fig3}), the surface brightness difference between the arc and the other regions in the inner hole has a 6$\sigma$ significance, whereas the other directions show small fluctuations in emission.  Furthermore, the polarization vectors in the region of the arc mostly face the central star as well as the region of the bright disk, indicating that the arc is also illuminated by central star and is physically connected to inner edge of the disk. 
  
In the same way, the dip D is a real dip feature and is not an artifact because the difference in surface brightness between D and the rest of the disk is statistically significant and not within errorbars.  In addition, the polarization vectors in the region of the dip mostly face the central star.

 There appears to be another dip located around a position angle $\sim$255$^{\circ}$, which is not azimuthally symmetric with D in terms of location and surface brightness.  However, this another dip is not significant outside of the uncertainties of the observations, thus follow up future observations are required to confirm its presence.


\section{Discussion}

\subsection{Comparison with the SMA Imaging} 
\label{sect:comp_SMA}

\citet{2012ApJ...753...59M} present SMA 880~$\mu$m continuum and  CO
(3-2) line observations of this  
source with a resolution of $0''.34$ and resolve the disk (Figure 1(d)).
They determine the surface density distribution of the disk 
with inner cavity whose radius is 72 AU based on the model 
fitting to the visibility at 880~$\mu$m. 
They also suggest that sparse matter still remains inside the cavity (20~-~72~AU) because of the dust emission indicated by mid-infrared \citep{2009ApJ...705...1646} excesses and their model fitting.
The NIR peak disk radius of 63~AU derived from our $H$-band image roughly agrees or may be slightly smaller than their measurement and can be explained by scattered light reflected at the inner edge of the disk. 


Based on their CO peak velocity map, \citet{2012ApJ...753...59M}
 derive the position angles of
the major and minor axes to be $85^{\circ}\pm10^{\circ}$ and
$-5^{\circ}\pm10^{\circ}$, respectively. 
Their derive disk
parameters, including the position angle  
and inclination, are in a good agreement with values derived from our near
infrared observations. 

  \citet{2012ApJ...753...59M} point out that the relatively high mass
  of the gas-dust disk,  
  its sharp inner edge, and the lack of gas accretion onto the star together 
  point to a dynamical origin for the inner hole. No companions have been found 
  down to brown dwarf masses \citep{2008ApJ...679..762, 2011ApJ...726..113}.  
  Based on their analysis, they suggest that the origin of 
  the hole can be attributed to the unseen presence of one or
  several giant planets within about a 40~AU.

The width and depth of the hole depend on the planet mass, disk
viscosity, and disk thickness 
\citep[e.g.,][]{1996ApJ...460..832T,2006Icar..181..587C}.  
As discussed by Dong et al.~(2012)b, the radius of the inner hole is on the
order of $\sim$~60~-~70~AU.  There have been extensive studies on the process of gap opening by a planet embedded in a disk
\citep[e.g.,][]{1986ApJ...307..395L,1996ApJ...460..832T,2006Icar..181..587C}.  
A rough estimate of the upper limit of the size of the gap 
that one planet can make may be given by the location of 
the lowest order Lindblad resonances, which, 
in the case of the outer resonance (the outer edge of the gap), is at
$\sim$1.6 times the orbital radius of 
the embedded planet \citep{1996ApJ...460..832T}.  If this is the case,
the existence of the inner hole might 
indicate the existence of at least one planet at $\sim$40~-~50~AU from the
central star \citep[see also][]{2006Icar..181..587C}, 
and the inner edge of the hole would be expected at $\sim 25-30$ AU, 
which is inside of our saturation radius and is therefore unresolved in this study.

If multiple planets reside within the hole, the
hole width can be wider and the inner disk may be much smaller than the
single planet case \citep{2011ApJ...729...47Z}.  To understand the 
properties of the whole of this object, it is important
to give constraints on the inner disk structure using 
detailed SED modeling (see Dong et al.~(2012)b) and/or high-resolution
observations by ALMA.  
Our observations do not show a clear inner disk edge, suggesting that the inner disk radius must be inside of the saturation radius of our observations.


\subsection{Non-axisymmetric Features}

 We list a few possibilities for origin of the two non-axisymmetric features, the dip and the arc since it is difficult to draw firm conclusions at this stage.
According to Figure 6.(c) in \citet{2012ApJ...753...59M}, the only
structures that are not reproduced by the axisymmetric disk model
within 2$\sigma$ error are the south-east component; the north dip is
consistent with the axisymmetric disk model convolved with the
elliptical beam.  We, therefore, do not elaborate on the north dip on
the sub-mm disk in the following discussions.
\citet{2012ApJ...753...59M} observe that 
the millimeter continuum emission has a peak at $0''.5$
southeast from the star in the 880~$\mu$m image.  
A similar distribution is also seen in the total integrated intensity distribution of CO (3-2) emission.  Our $H$-band image and Figure \ref{fig4}
show that the southeast of the disk is brighter than the east and the west, 
despite the fact that we
observe the scattered light originating at the disk surface.    
This may be explained in a way that the more materials exist, the
thicker the disk is, 
resulting in a higher grazing angle of the stellar light at the
scattering surface \citep[e.g.,][]{2011ApJ...739...10M}.
In our near-infrared image, the north of the disk is also bright.  
The brightness of the disk in the north, together with sub-mm observations suggesting material accumulation in the south, infers that the northern portion of
the disk is the near side. This assumes that forward scattering dominates.
If this picture of the disk orientation is correct,
the disk rotation is counter-clockwise direction.
This geometry might also explain the 
origins of the non-axisymmetric structures at D.
 D reside at the interface between where the material 
accumulated (south) and where forward scattering is significant
(north).
However, it should be noted that since the disk is nearly face-on,
significant forward scattering excess and precise prescriptions of the 
phase function should be required, and our discussion crucially depends
on the details of the optical properties of the dust particles.  
It should also be noted that the material concentration in the south is
within 2$\sigma$ of the current sub-mm data \citep{2012ApJ...753...59M}.  The
detection of the non-axisymmetric structures at sub-mm wavelengths is needed to examine the details of the disk geometry.

Another possibility for the origin of the dip D is that there is a massive gravitating
object at the disk's midplane.  It could be a planet (which is not
necessarily the same as the one discussed in the previous section as the
origin of the inner hole), a clump formed by hydrodynamic
instability (e.g., gravitational instability), or something else.  Such
objects can pull the materials from the surface toward the disk's midplane,
thereby producing a shadow in the scattered light image\citep{2009ApJ...700..820J}.
This might produce features such as D.  
Such objects have sizes smaller than the SMA beam size, therefore cannot be resolved with the SMA.

For the arc structure in the west, the morphology is similar to spiral density waves in the disk.  An arguably uniqueness is that this arc extends over the ``disk inner hole'' region.  It also resembles to spiral density waves in terms of the pitch angle.  Spiral waves can be produced by dynamical processes in the disk such as the turbulence or
the planet-disk interaction \citep[see e.g.,][]{2007prpl.conf..655P}. 
 If this feature is a spiral wave, its
morphology indicates that the disk rotation is counter-clockwise because
spirals are in general trailing features.  This rotation
direction is consistent with the picture presented at the beginning of
this section. 
Last, we note that the arc
structure shows morphological similarity to the  
``planet shadows'' discussed by \citet{2009ApJ...700..820J}. 
One possible way to distinguish the different scenarios is 
to observe the structures of the disk's midplane 
at longer wavelength.  
For instance, if a density enhancement is observed at the location of D, this might indicate the
existence of a massive body there.  If we see an arc structure in the
sub-mm as well, then it is likely a density wave.

 Another possible way to reveal the nature of the arc may be the
contemporaneous SED and scattered light observations.  Previous studies indicate that substantial variability has also been seen in MIR SED and has been linked to variable structures including arcs or rims of the disk \citep{2009AJ....137.4024D,2008ApJ...682..548W,2009ApJ...704L..15M,2011ApJ...728...49E,2012ApJ...748...71F}.  For J1604-2130 case, the location of the arc and the rim of the disk wall is generally consistent with the radial location in the disk which possibly contributes to the SED variability.  


\acknowledgments

We thank the telescope staff and operators at the Subaru Telescope for their assistance.  We also thank our referee for the constructive comments that have helped to improve this manuscript.  Part of this research was carried out at the JPL, under a contract with the NASA.
This work is partially supported by a Grant-in-Aid for Science Research in a Priority Area from MEXT 22000005(M. T.), KAKENHI 23103004(M. M. and M. F.), 24103504(T. T.), 24840037(T. M.), US NSF grant 1009314(J. P. W.), and 1009203(J. C).  This work was supported in part by the Center for the Promotion of Integrated Sciences (CPIS) of The Graduate University for Advanced Studies~(SOKENDAI).

\clearpage



\begin{figure}
\includegraphics[angle=0,scale=1.20]{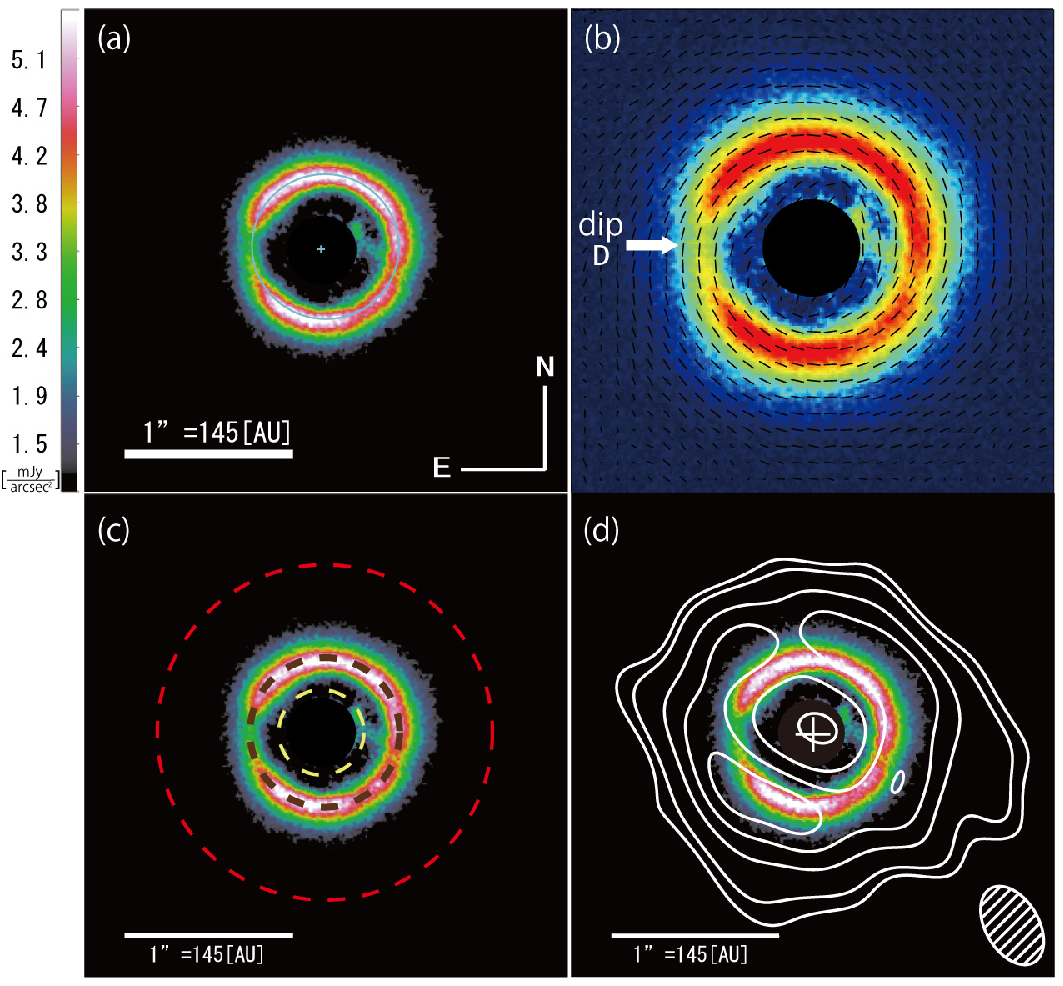}
\caption{$H$-band HiCIAO images of J1604-2130.  
The saturated central area
 (radius = 0.$''$2) is masked in black.  
 (a): The $PI$ image of J1604-2130.  The field of
 view~(FOV) is $2''.9 \times 2''.9$.  
The unit of the color bar is mJy/arcsec$^2$.  
 The light blue ellipse and plus sign are the best
 fit result of our elliptical disk model and the ellipse center.
 (b): $H$-band polarization vectors superposed on the $PI$ image.  
The vector directions indicate angles of
polarization. The plotted vectors are based on 7 [pixel]
 $\times$ 7 [pixel] binning corresponding to the spatial resolution.
 The FOV is $2''.0 \times 2''.0$.  The vector's lengths are
 arbitrary. 
  (c):Red($r$= 145~AU), brown($r$ = 63~AU), and yellow($r$ = 33~AU) circles, corresponding figure~\ref{fig3}, superimposed on the $PI$ image.
 (d):SMA 880~$\mu$m continuum map \citep{2012ApJ...753...59M}
 superimposed on the $PI$ image.  White color
 contours indicate 2, 3, 6, 9, and 12$\sigma$ intensity (1$\sigma$ =
 1.3 mJy/beam).  The $\sim 0.''5 \times0.''3$ beam of SMA is shown in the
 bottom right.
 }
.\label{fig1}
\end{figure}


\begin{figure}
\includegraphics[angle=0,scale=.90]{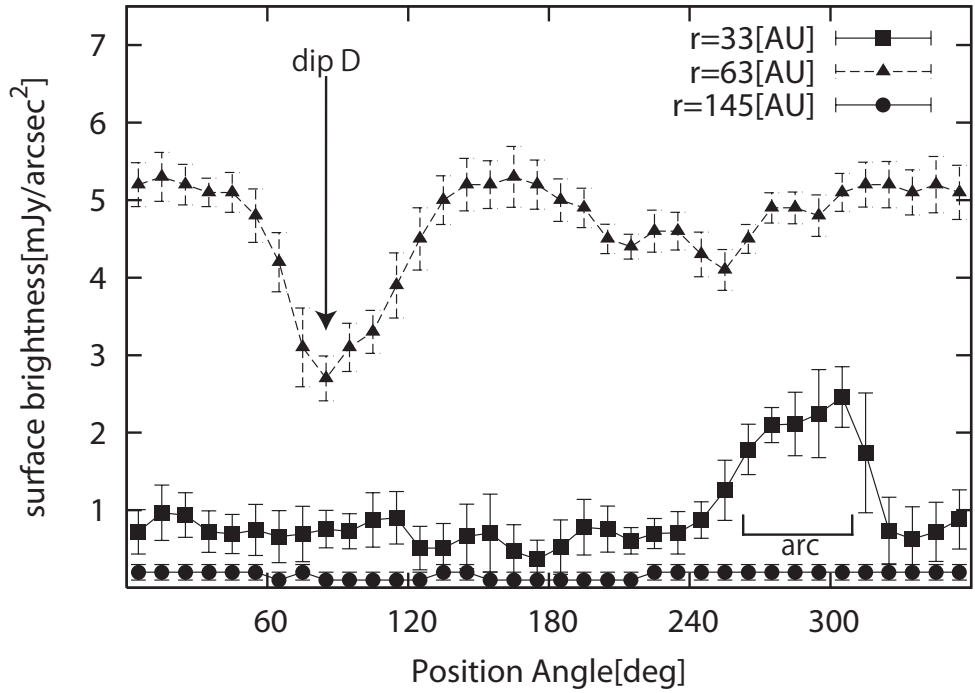}
\caption{ Azimuthal surface brightness profile at $r$ = 33~AU (inner hole region), $r$ = 63~AU (inner disk region), and $r$= 145~AU (outer region) with position angle measured from north to east.  The radius 33 AU is measured as it is the brightest part of the arc.
}
.\label{fig3}
\end{figure}

\begin{figure}
\includegraphics[angle=0,scale=.90]{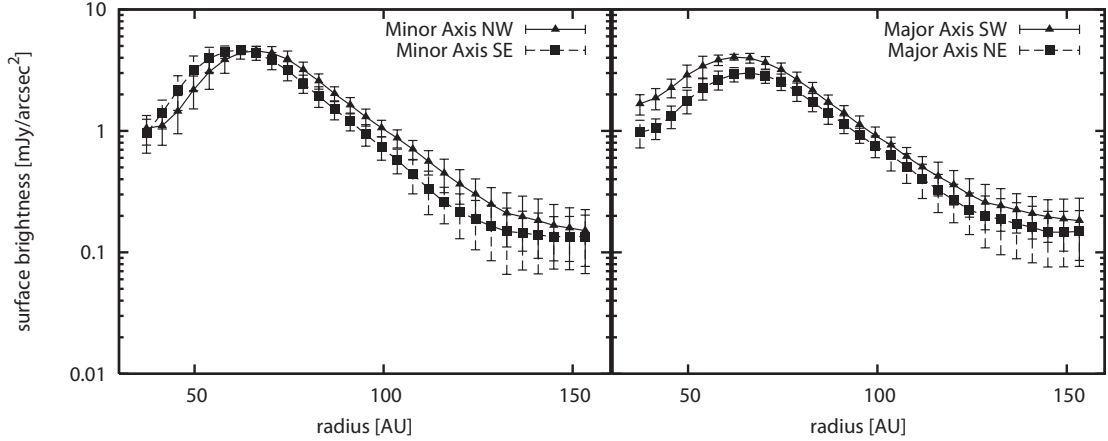}
\caption{Left Panel:Radial profile along the minor axis.  NW and SE radial profile are averaged over -25$^{\circ}$~\textless P.A. \textless ~5$^{\circ}$
 and 155$^{\circ}$~\textless P.A. \textless ~185$^{\circ}$,
 respectively.  This is because the disk position angle derived from
 our near-infrared image and peak velocity of the submm CO emission ($-5^{\circ}\pm10^{\circ}$) are 9$^{\circ}$
 different.  Therefore, we centered our 30$^{\circ}$ wide radial profiles on the average of our near-infrared and submm line emission.
 Right Panel:Radial profile along the major axis.  In the same way, NE and SW radial profiles
 are averaged over 65$^{\circ}$~\textless P.A. \textless ~95$^{\circ}$
 and 245$^{\circ}$~\textless P.A. \textless ~275$^{\circ}$,
 respectively.  
}
.\label{fig4}
\end{figure}

\clearpage

\begin{table}
\begin{center}
\caption{The result of the ellipse fitting for the disk of J1604-2130}
\begin{tabular}{lr}
\tableline
Semimajor axis &$63.4 \pm 1.0$ [AU]\\
Semiminor axis &$62.4\pm 1.0$ [AU]\\
Ellipse center (u,v)\tablenotemark{a}&($-1.1\pm1.1,1.5\pm1.1)$ [pixel]\\
Position angle &$-14\pm11^{\circ}$\\
Inclination & $10.2\pm10.1^{\circ}$\\
\tableline
\end{tabular}
\tablenotetext{a}{Central position (0, 0) corresponds to the stellar position.}
\end{center}
\end{table}

\clearpage

\end{document}